\documentclass[
    ,final            
  ]
  {aipproc}

\layoutstyle{6x9}

\begin{document}

\title{Search for Physics beyond the Standard Model\\
at the Tevatron}


\author{Jean-Fran\c{c}ois Grivaz\\
{\small on behalf of the CDF and D\O\ Collaborations}}{
  address={Laboratoire de l'Acc\'el\'erateur Lin\'eaire -- IN2P3-CNRS \& Universit\'e de Paris-Sud-XI\\
Centre Scientifique d'Orsay -- B\^at. 200 -- BP 34 -- 91898 Orsay Cedex -- France}
}

\begin{abstract}
Recent searches for physics beyond the standard model at the Tevatron are 
reported, with emphasis on supersymmetry.
\end{abstract}

\maketitle

\section{Introduction}

This talk is devoted to searches for signals of physics beyond the standard 
model, performed by the CDF and D\O\ experiments at the Fermilab Tevatron, 
where protons and antiprotons collide at a center of mass energy of 1.96~TeV. 
The results reported here are based on data samples corresponding to integrated
luminosities of up to 1.2~fb$^{-1}$. Details can be found in Ref.~\cite{url}.

The standard model (SM) is constructed with the following ingredients: a field 
theory in a four-dimensional space time, with invariance under the Poincar\'e 
group; the SU(3)$_{\mathrm C}\times$SU(2)$_{\mathrm L}\times$U(1)$_{\mathrm Y}$
gauge group, with electroweak symmetry breaking (EWSB) by the Higgs mechanism; 
three families of quarks and leptons. 

Possible extensions of the SM include: extending the Poincar\'e group by 
supersymmetry (SUSY) which, in its local form, allows gravitation to 
be incorporated in the theory; replacing the field theory by a 
(super)string theory; increasing the number of space dimensions; embedding the 
SM gauge group in a larger one, in the framework of grand unified theories 
(GUTs); introducing new interactions between quarks and leptons, mediated by
leptoquarks; invoking 
alternative mechanisms of EWSB, such as technicolor; attempting to repeat the 
history of onion peeling in theories of compositeness... 

Because of the limited
duration of this talk and of the very nature of this conference, I will 
concentrate on SUSY and extra dimensions, after a brief excursion into 
GUT motivated searches.
All limits will be given at the 95\% C.L., and discoveries will be reported
only for a level of significance of at least $5\sigma$.

\section{Additional gauge bosons}

\begin{figure}
\begin{tabular}{cc}
  \includegraphics[height=.3\textheight]{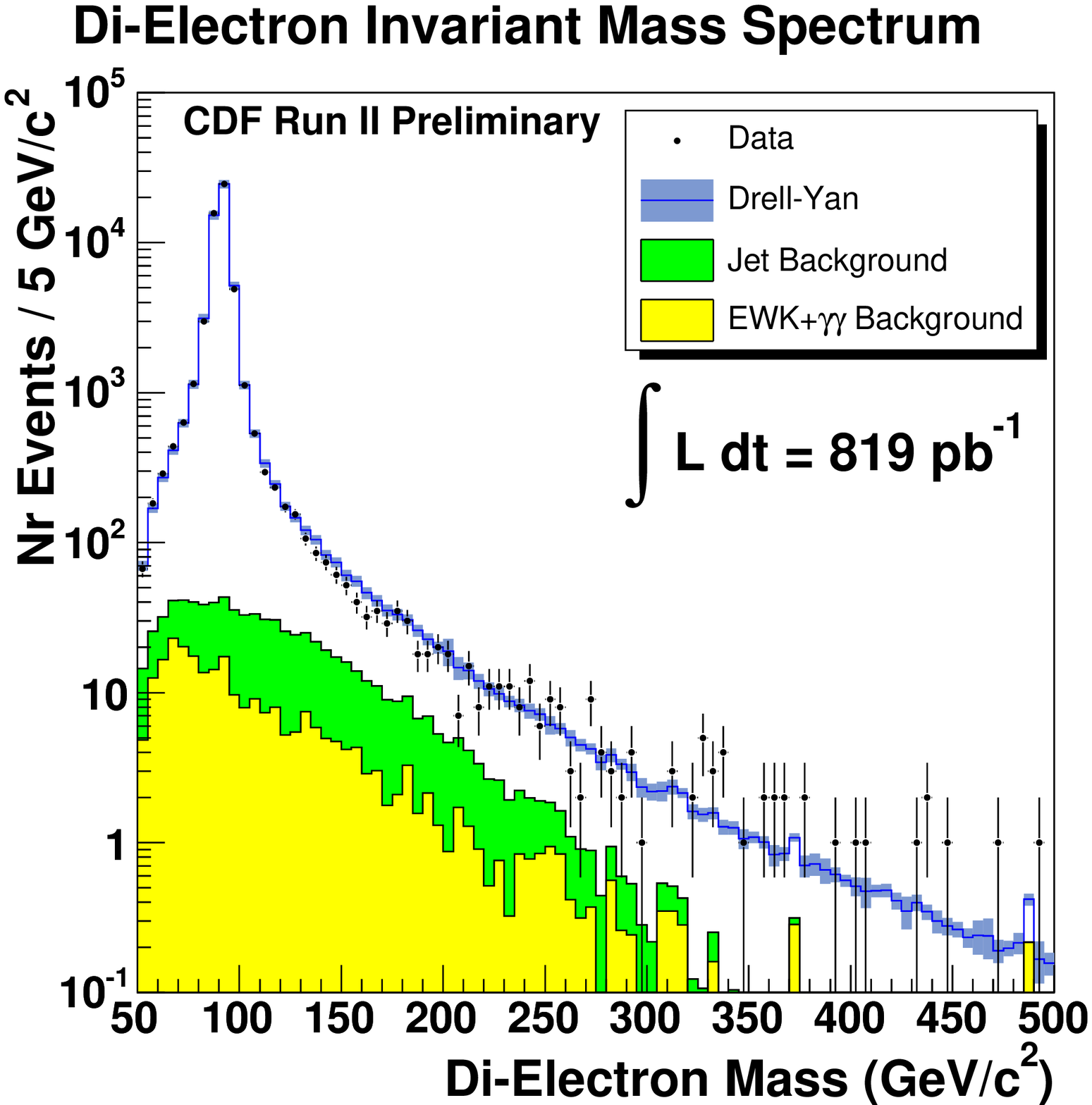} &
  \includegraphics[height=.3\textheight]{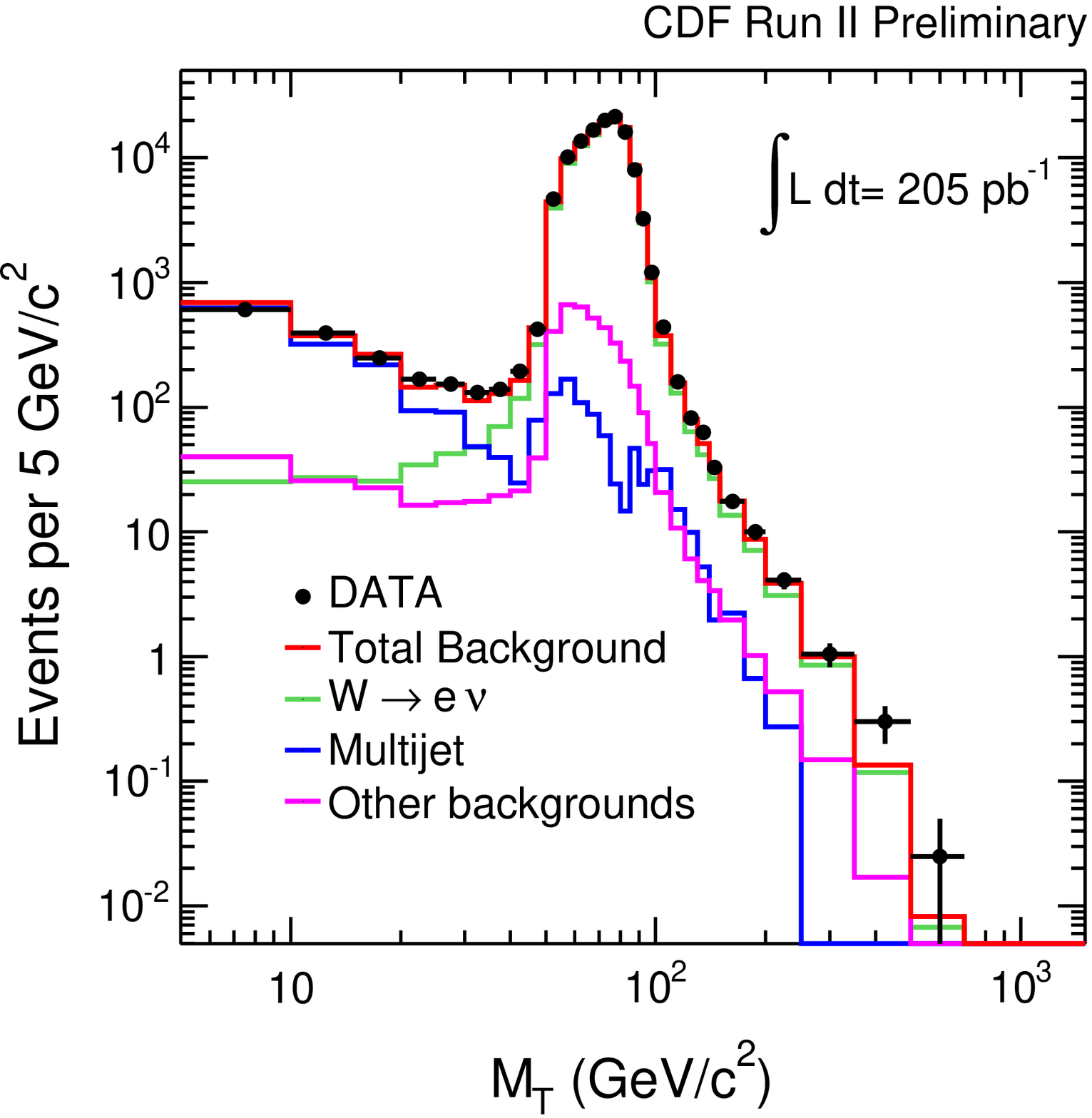}
  \caption{Dielectron mass (left) and electron-missing $E_T$ transverse mass 
(right).}
\end{tabular}
  \label{prime}
\end{figure}

Additional neutral bosons similar to the $Z$ arise for instance in string 
inspired GUTs where extra U(1)s result from the breaking of an E(6) group.
Such $Z'$ bosons can be produced by the Drell-Yan mechanism, and are 
best searched via their decay into an $e^+e^-$ pair because of the low 
background, compared to $q\bar q$ decays, and of the superior mass resolution, 
compared to $\mu^+\mu^-$ decays. The absence of any peak (other than the $Z$) 
in the dielectron mass spectrum observed by CDF in 0.8~fb$^{-1}$ 
(Fig.~\ref{prime}) allows a lower limit of 850~GeV
to be set on a ``sequential'' $Z'$, i.e. with couplings identical to those of 
the SM $Z$ boson~\cite{phb}. 

Similarly, extra $W$ bosons are expected in left-right symmetric models. The 
search is performed in the $W'\to e\nu$ channel, where the dielectron mass is 
replaced by the transverse mass, constructed from the transverse momentum of an
isolated electron and the missing transverse energy. No Jacobian peak other 
than the one from the $W$ is seen (Fig.~\ref{prime}), which yields a $W'$ lower
mass limit of 788~GeV, based on 0.2~fb$^{-1}$ of CDF data. 

\section{Supersymmetry}

In such a conference, it is probably not appropriate to address questions such 
as ``What is SUSY ?'' or ``Why SUSY ?''. Nor to define squarks or gauginos, or
$R$-parity\dots I will therefore restrict myself to:  ``Which SUSY ?''.

The Tevatron experiments have investigated various incarnations of SUSY:
\begin{itemize}
\item a more or less constrained MSSM (minimal supersymmetric extension of the
SM), the most constrained form being mSUGRA (minimal supergravity) controlled
by five parameters ($m_0$, $m_{1/2}$, $\tan\beta$, $A_0$ and sign($\mu$)). A 
neutralino LSP (lightest SUSY particle), $\tilde\chi^0_1$, is assumed. Both $R$-parity
conservation and violation are considered;
\item gauge-mediated SUSY breaking (GMSB), with a gravitino LSP and a 
neutralino NLSP (next to lightest SUSY particle);
\item anomaly-mediated SUSY breaking (AMSB), with a wino LSP and a long-lived 
chargino;
\item split-SUSY, with heavy scalars and a long-lived gluino.
\end{itemize}

\subsection{Searches in $R$-parity conserving supergravity inspired models}

Detailed analyses were performed in the mSUGRA framework, along two main 
directions:
\begin{itemize}
\item the search for squarks and gluinos, leading to multijet+missing $E_T$
topologies, with large production cross sections but also large instrumental
backgrounds;
\item the search for electroweak gauginos with leptonic decays, leading to the
celebrated trilepton+missing $E_T$ final state, with smaller production cross 
sections times leptonic branching fractions but with a clean experimental 
signature.
\end{itemize}

\paragraph{\bf Charginos and neutralinos~\cite{js,mh}}

Associated production of the lightest chargino (with $\tilde\chi^\pm\to\ell\nu\tilde\chi^0_1$) 
and second lightest neutralino (with $\tilde\chi^0_2\to\ell^+\ell^-\tilde\chi^0_1$) lead to 
final states containing three leptons and missing $E_T$ due to the neutrino 
and to the weakly interacting $\tilde\chi^0_1$s. In spite of its apparent simplicity, 
this search is challenging because the rates are low, the leptons are soft, 
and the $\tau$ contribution is enhanced at large $\tan\beta$. A large 
integrated luminosity and the combination of many final states are necessary
to reach a relevant level of sensitivity. 

The general strategy is to require two isolated electrons or muons and some 
missing $E_T$. Channel dependent selection criteria are applied to reject 
backgrounds, e.g. from $Z\to\ell\ell$. Finally, either the two leptons are 
required to have the same sign, a configuration in which the SM backgrounds are
small, or a third lepton is required, identified or in the form of an isolated 
track, the latter providing some sensitivity to hadronic $\tau$ decays. 

In the end, the remaining backgrounds are instrumental, due to lepton 
misidentification or to fake missing $E_T$, or irreducible from SM processes 
such as $WZ$ production. Altogether, the CDF analyses select 11 events while 
$9.0 \pm 1.0$ are expected. The corresponding numbers for D\O\ are 4 and 
$4.9 \pm 1.0$. Limits on the production cross section times branching fraction 
into three leptons were derived, with some model dependence as explained below.
For $\tilde\chi^\pm$ and $\tilde\chi^0_2$ masses of 
140~GeV, they are at the level of 0.22~pb for CDF and 0.07~pb for D\O. (The 
corresponding limits expected in the absence of a signal are 0.16 and 0.08~pb.)
   
\begin{figure}
\begin{tabular}{cc}
  \includegraphics[height=.3\textheight]{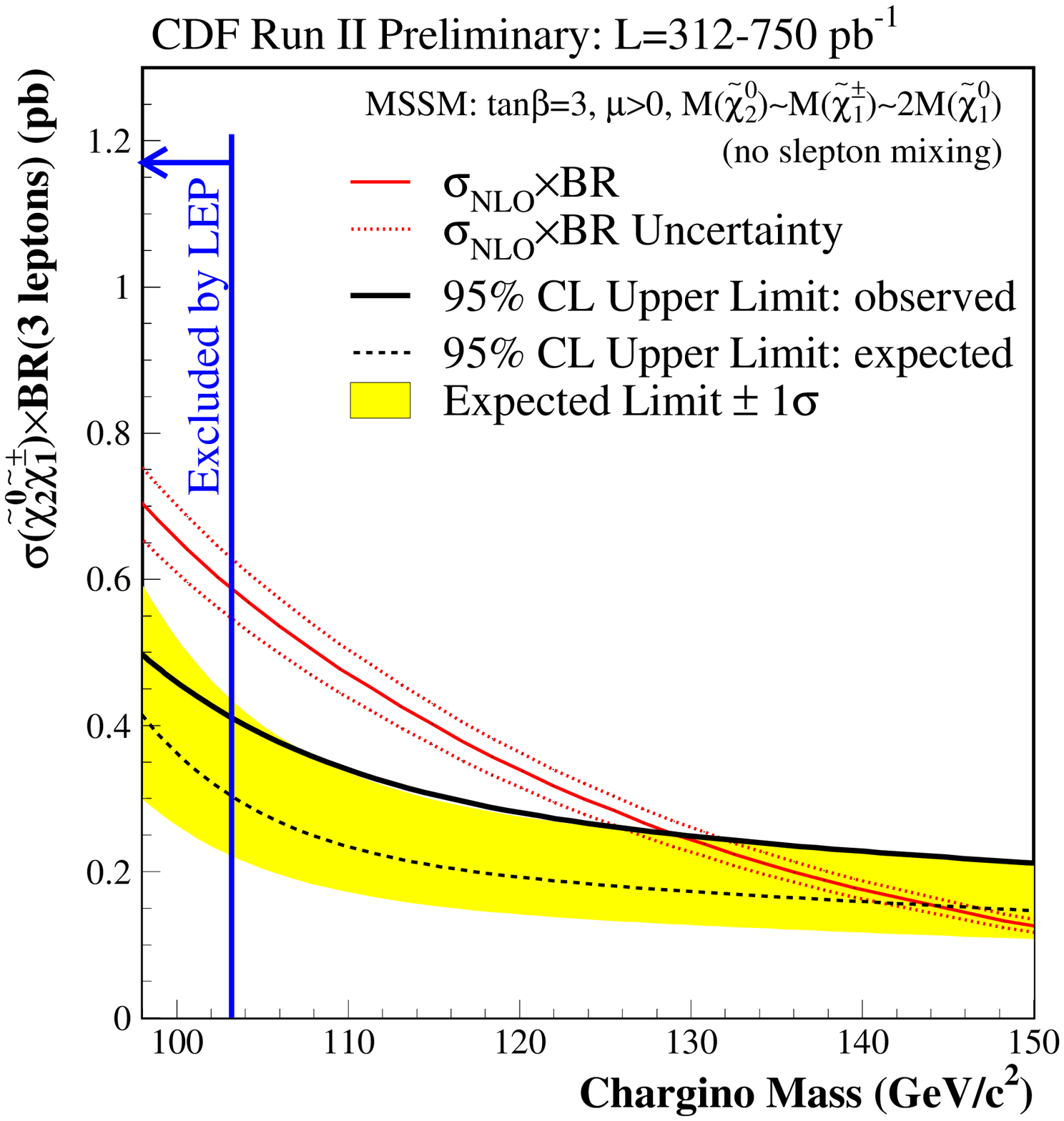} &
  \includegraphics[height=.25\textheight]{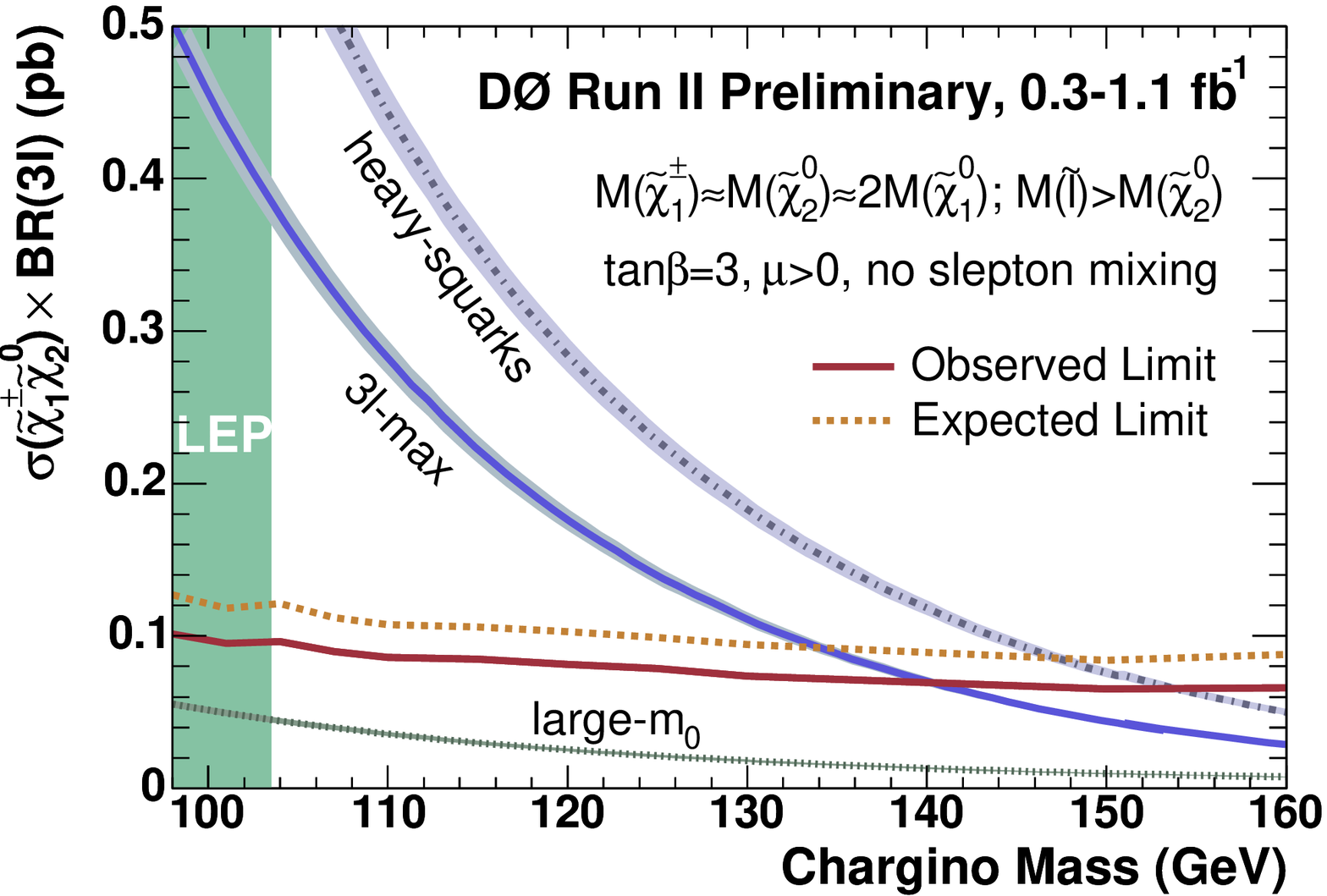}
  \caption{Limit on the associated $\tilde\chi^\pm\tilde\chi^0_2$ production cross section
times branching fraction into three leptons by CDF (left) and D\O\ (right). 
Theoretical expectations in various mSUGRA scenarios are also indicated.}
\end{tabular}
  \label{trilep}
\end{figure}

To turn these cross section limits into mass limits, specific mSUGRA 
configurations were considered. In the mass range of interest, the 
$\tilde\chi^\pm$ 
and $\tilde\chi^0_2$ decays proceed via virtual $W$ or $Z$ boson exchange, or via 
slepton exchange. If sleptons are heavy, the leptonic branching fractions are 
too small to allow any mass limit to be placed. Low slepton mass configurations
have been considered by both collaborations: with a value of $m_0$ fixed at 
60~GeV by CDF, with $m_0$ adjusted such that the slepton masses are just above
the $\tilde\chi^0_2$ mass by D\O\ (the ``$3\ell$-max'' scenario); in both cases, 
slepton masses were set to be equal for the three lepton flavors. The results 
are shown in 
Fig.~\ref{trilep}. The chargino mass lower limit obtained by D\O\ in the 
$3\ell$-max scenario is 146~GeV, well beyond the LEP limit of 103.5~GeV.
The CDF limit in the low $m_0$ scenario they considered is 126 GeV.

\paragraph{\bf Generic squarks and gluinos~\cite{pb,mm}}
    
If squarks are lighter than gluinos, they are expected to decay according to 
$\tilde q\to q\tilde\chi^0_1$, while gluinos are expected to decay according to 
$\tilde g\to q\bar q\tilde\chi^0_1$ if squarks are heavier than gluinos. Jets and 
missing $E_T$ carried away by the final $\tilde\chi^0_1$s is therefore expected from any 
squark or gluino production. The pair production of light squarks leads to 
(at least) two jets; the pair production of light gluinos leads to (at least) 
four jets; the associated production of a squark and a gluino of similar masses
leads to (at least) three jets. The reason for the ``(at least)s'' is that 
initial or final state radiation tends to increase the jet multiplicity. The
possibility of cascade decays such as $\tilde q\to q'\tilde\chi^\pm$ complicates the
picture, and a specific model such as mSUGRA is therefore needed to interpret 
the search results.

The main backgrounds in a search for multijets with missing $E_T$ are
\begin{itemize} 
\item instrumental, from QCD multijet production with fake missing $E_T$ due to
jet energy mismeasurements,
\item from the associated production of $W$ and jets, with $W\to\ell\nu$ where 
the lepton is not identified,
\item from the associated production of $Z$ and jets, with $Z\to\nu\nu$, which
constitutes an irreducible background.
\end{itemize}

The CDF analysis was optimized for the configuration where the squark and 
gluino masses are similar, while D\O\ developed three analyses optimized for
different hierarchies of squark and gluino masses. The main selection criteria
are for each analysis a minimum number of jets, a minimum missing $E_T$, a
minimum value for $H_T$, the sum of jet transverse energies, a veto on isolated
leptons, and topological cuts on the angles between the missing $E_T$ and jet 
directions. In 370~pb$^{-1}$, the CDF analysis selected two events for a 
background expectation of $8.2 \pm 2.9$ events. The three D\O\ analyses 
selected 18 events altogether in 310~pb$^{-1}$,
for a total background expectation of $19.0 \pm 5.3$ events.

\begin{figure}
\begin{tabular}{cc}
  \includegraphics[height=.3\textheight]{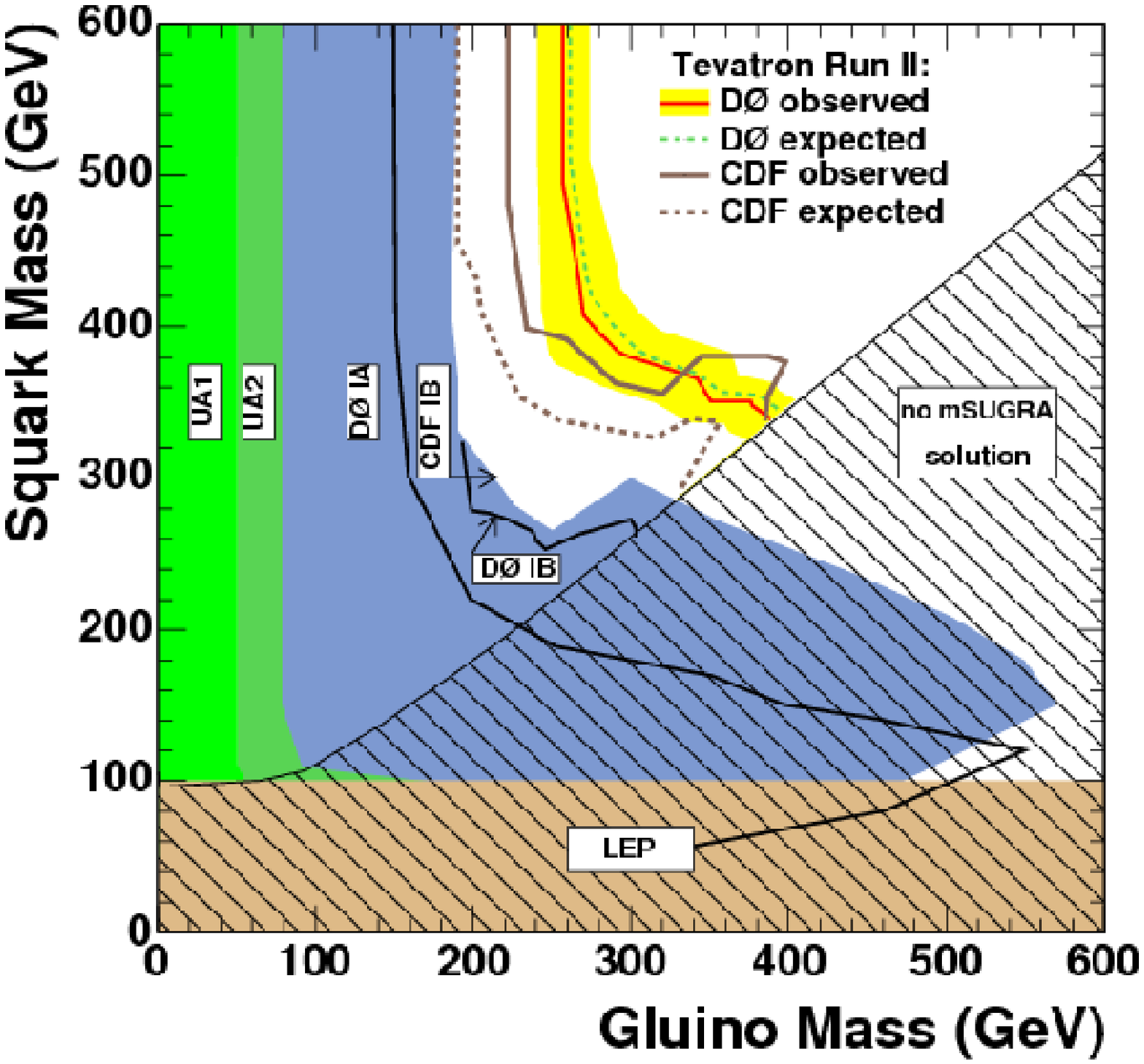} &
  \includegraphics[height=.3\textheight]{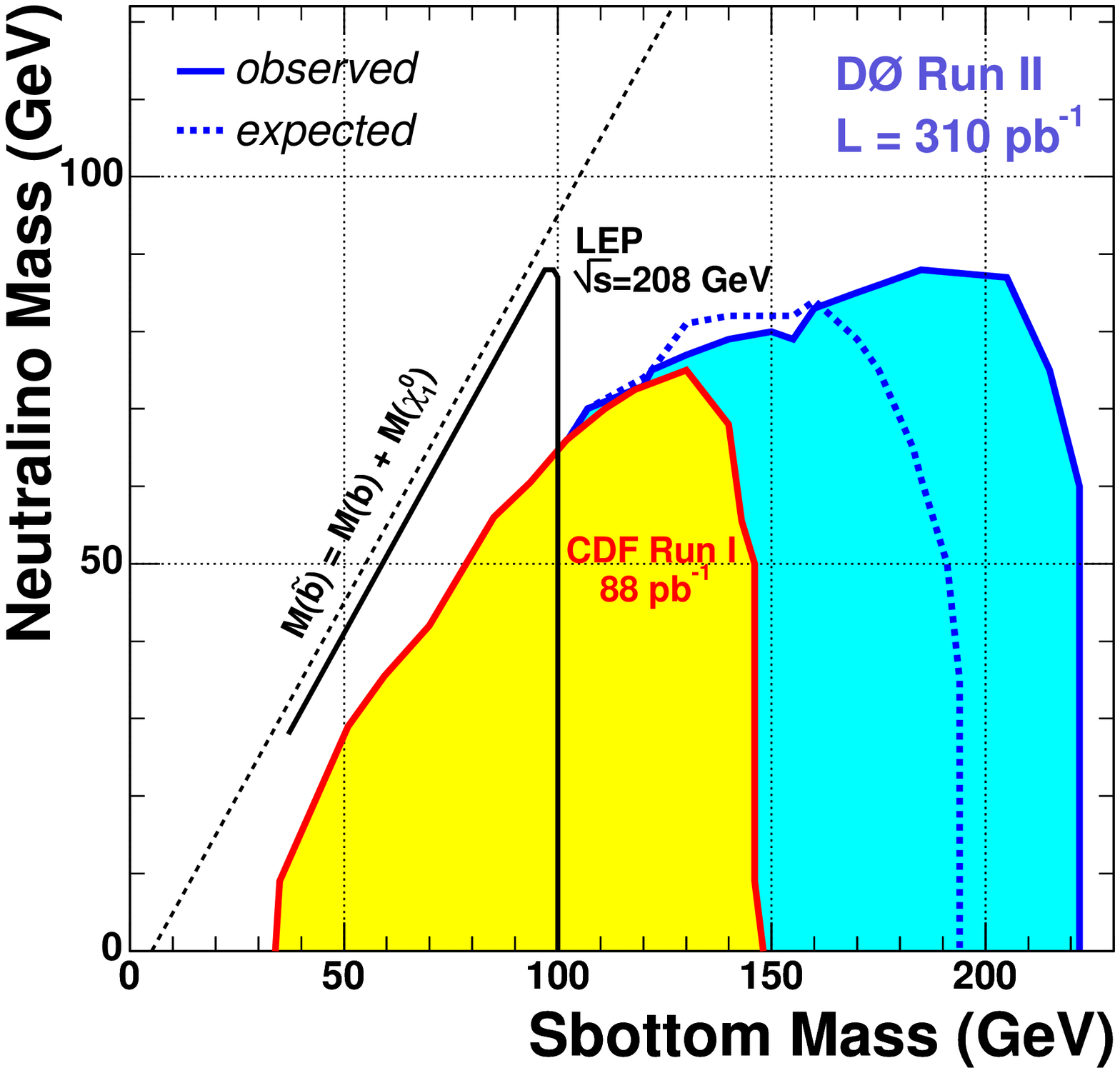}
  \caption{Excluded domains in the plane of generic squark and gluino masses 
(left), and in the plane of the sbottom and $\tilde\chi^0_1$ masses (right).}
\end{tabular}
  \label{sqgl}
\end{figure}

To turn these results into exclusion domains in the plane of the squark and 
gluino masses (Fig.~\ref{sqgl}), the mSUGRA framework was used, with mass 
degenerate light-flavor squarks. The results from the two collaborations are
not directly comparable because D\O\ handles the theoretical uncertainties on
the production cross sections in a somewhat more conservative way than CDF
(lower edge of the light shaded band in Fig.~\ref{sqgl}).
Taking the CDF and D\O\ results as quoted by the authors, mass lower limits of 
241 and 325~GeV are obtained for gluinos and squarks, respectively, and of 
387~GeV for equal-mass squarks and gluinos. 

\paragraph{\bf Third generation squarks~\cite{pb,mm}}

Because of the impact of the large top Yukawa coupling on the renormalization 
group equations, the stop squarks can be expected to be lighter than generic
squarks. Furthermore, the mixing in the stop mass matrix is also enhanced by
the large top quark mass, so that the lighter stop could well be the lightest
of all squarks, which justifies dedicated searches. The D\O\ collaboration has
considered two scenarios
\begin{itemize}
\item a stop NLSP with a $\tilde\chi^0_1$ LSP, in which case the decay mode is the 
flavor changing process $\tilde t\to c\tilde\chi^0_1$,
\item a mass hierarchy involving a light sneutrino, so that the stop decays
according to $\tilde t\to b\ell\tilde\nu$. 
(The subsequent $\tilde\nu\to\nu\tilde\chi^0_1$ decay involves only invisible particles.)
\end{itemize}

\begin{figure}
\begin{tabular}{cc}
  \includegraphics[height=.3\textheight]{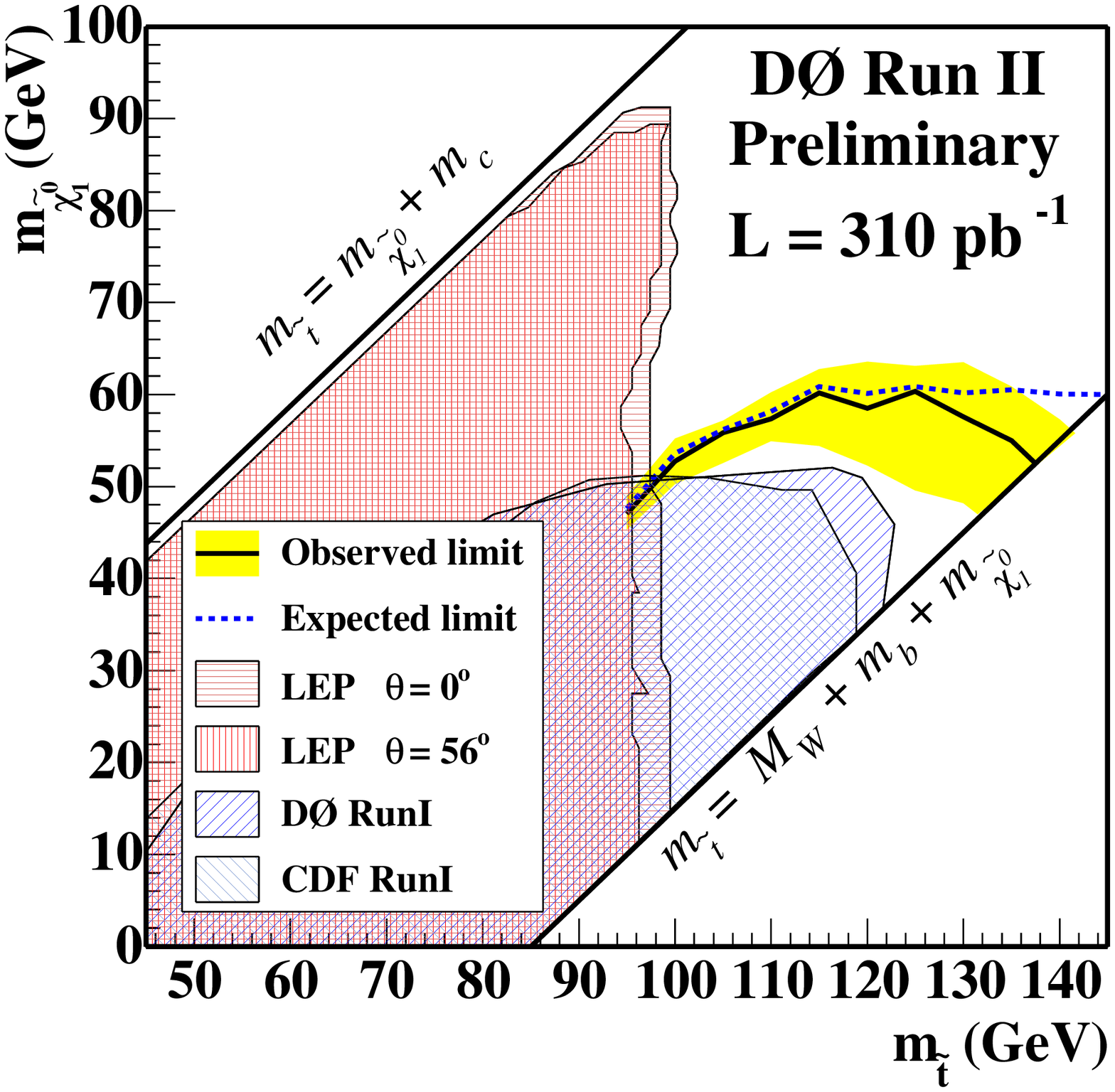} &
  \includegraphics[height=.3\textheight]{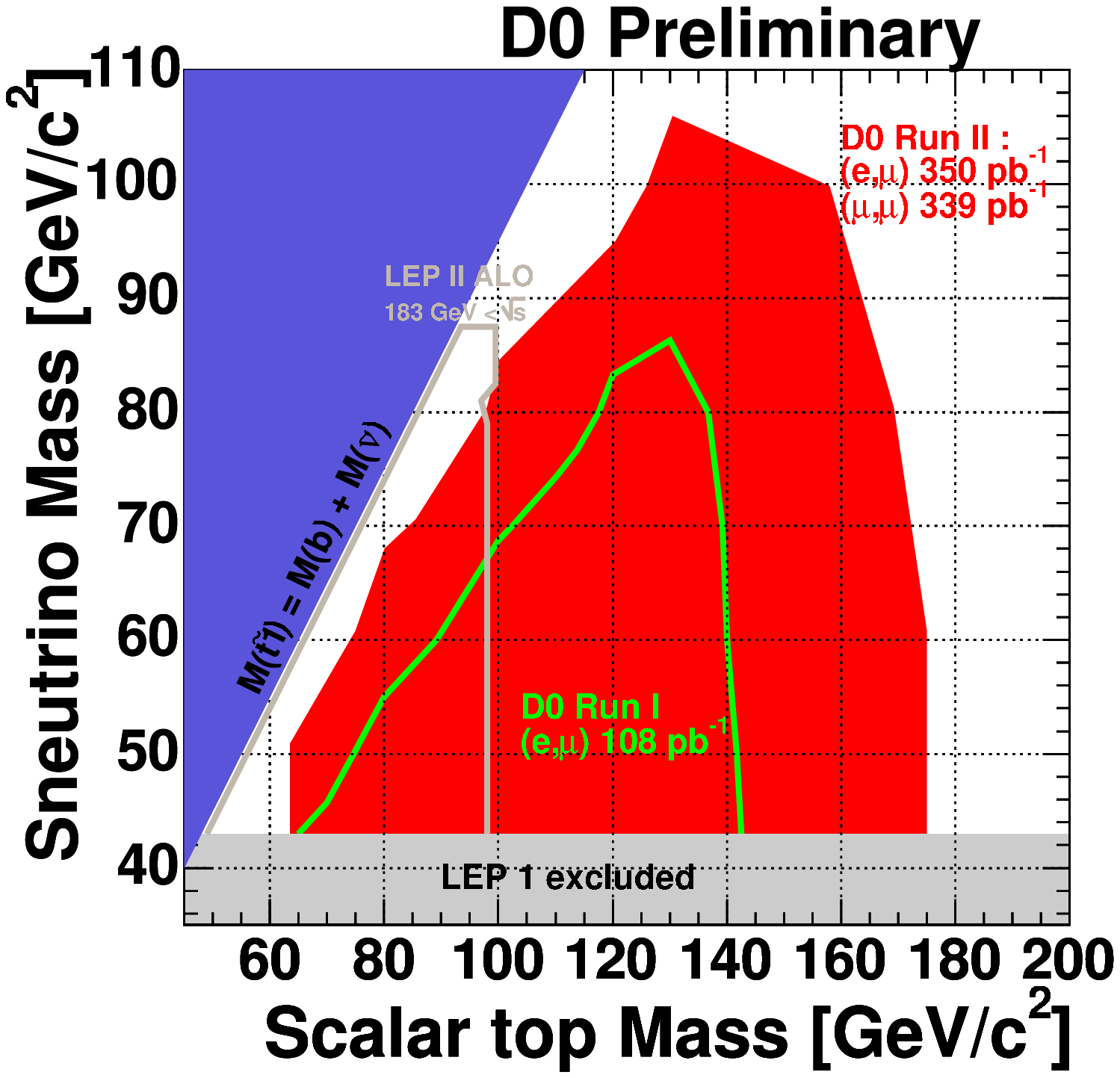}
  \caption{Excluded domains in the plane of the stop and $\tilde\chi^0_1$ (left), and
stop and sneutrino (right) masses.} 
\end{tabular}
  \label{stop}
\end{figure}

In the first scenario, stop pair production leads to an acoplanar jet final 
state similar to the one considered for generic squarks. The difference is
that, in the mass range of interest, the jets are much softer, and therefore 
the QCD background larger. To overcome this difficulty, a soft heavy-flavor
jet tagging is applied. The kinematic selection is optimized as a function of
the stop and $\tilde\chi^0_1$ masses. For $m_{\tilde t}=130$~GeV and $m_{\tilde\chi^0_1}=50$~GeV, 60
events are selected while the SM background expectation is $59.4 \pm 12.8$
events. An excluded domain in the plane of the stop and $\tilde\chi^0_1$ masses is 
derived (Fig.~\ref{stop}).

In the second scenario, the final state consists of two leptons, two $b$ jets 
and
missing $E_T$. Two analyses were developed, one for the $e\mu$ channel and the
other for the dimuon channel. Again, the kinematic cuts were optimized as a
function of the stop and sneutrino masses. As an example, the number of events
observed in the $e\mu$ channel for small $\tilde t$--$\tilde\chi^0_1$ mass differences is
21, while the expected background is $23.0 \pm 3.1$. The two channels were 
combined to yield the excluded domain in the plane of the stop and sneutrino
masses shown in Fig.~\ref{stop}.

Mixing can also be enhanced in the sbottom sector for large values of
$\tan\beta$. A search for sbottom pair production was performed by D\O, where
the $\tilde b\to b\tilde\chi^0_1$ decay is assumed. The analysis is similar to the one
applied for the stop search in acoplanar jets, except that a tight $b$ tagging 
can be applied, which provides sensitivity to higher masses, as can be seen in 
the exclusion domain shown in Fig.~\ref{sqgl}. The CDF collaboration considered
the mass hierarchy such that the gluino decays into $b\tilde b$. Gluino pair
production therefore leads to final states consisting of four $b$ jets and 
missing $E_T$. With 156~pb$^{-1}$, sbottom masses up to 240~GeV have been 
excluded in this specific configuration for gluino masses smaller than 280~GeV 
and for $m_{\tilde\chi^0_1}=60$~GeV.

\subsection{Supersymmetry with $R$-parity violation}

The most general superpotential contains
$$W_\mathrm{RPV}=\lambda_{ijk}L_iL_j\bar E_k+\lambda'_{ijk}L_iQ_j\bar D_k+
\lambda''_{ijk}\bar U_i\bar D_j\bar D_k$$
where $L$ and $Q$ are lepton and quark doublet superfields, $E$, $U$ and $D$ 
are lepton and quark singlet superfields, and $i,j,k$ are generation indices.
These lepton or baryon number violating terms are forbidden by $R$-parity 
conservation, but could be present provided they do not introduce unacceptably 
fast proton decay or flavor changing neutral currents. This is easily achieved 
if all but one of the trilinear couplings can be neglected. The CDF and D\O\ 
collaborations have considered a number of scenarios where this constraint is 
satisfied.

Pair production of gauginos was investigated by CDF and D\O, where the produced
SUSY particle decays lead to two $\tilde\chi^0_1$ LSPs. With a $\lambda_{121}$ coupling,
for instance, the $\tilde\chi^0_1$s further decay into $ee\nu_\mu$ or $e\mu\nu_e$ and the
final state therefore contains four electrons, three electrons and a muon, or 
two electrons and two muons, all with missing $E_T$ taken away by two 
neutrinos. Searches for three leptons (electrons or muons) allowed the 
$\lambda_{121}$ and $\lambda_{122}$ couplings to be probed, and a dedicated 
search by D\O\ for two electrons and a hadronically decaying $\tau$ increased
the sensitivity to a $\lambda_{133}$ coupling~\cite{ca}. 
Within the mSUGRA framework and 
for $\tan\beta=5$ and large slepton masses, lower mass limits of about 120~GeV
(105~GeV) 
were obtained by D\O\ (CDF) in the first two cases, using $\sim 350$~pb$^{-1}$.
For a $\lambda_{133}$ coupling, 
the search is most sensitive at large $\tan\beta$ and for light staus: a $\tilde\chi^0_1$
mass limit of 115~GeV was obtained by D\O\ 
for $\tan\beta=20$ and $m_0=100$~GeV. 

With a $\lambda'_{211}$ coupling, resonant smuon or sneutrino
production can occur via $q\bar q$ annihilation. In the case of smuon 
production, for instance, the smuon decays into a muon and a $\tilde\chi^0_1$, which in
turn can decay through the same $R$-parity violating coupling into a muon and 
two 
quarks. The final state products, two muons and two jets, allow the $\tilde\chi^0_1$ and
smuon masses to be reconstructed. The absence of accumulations for such masses 
in 380~pb$^{-1}$ 
allowed D\O\ to set limits on the cross section for a resonantly produced smuon
or sneutrino~\cite{ca}. 
Since the production cross section depends on the value of the
$\lambda'_{211}$ coupling, mass lower limits were obtained as a function of 
this coupling. They range from 210 to 363~GeV for $\lambda'_{211}=0.04$ to 
$0.10$.

In the case of a $\lambda'_{333}$ coupling, the $\tilde t\to b\tau$ decay is
expected to occur. Stop pair production therefore leads to a final state 
containing two $b$ jets and two $\tau$s. The CDF search for this topology 
selected two events in 320~pb$^{-1}$
for an expected background of $2.3 \pm 0.5$ events, which
lead to a lower stop mass limit of 155~GeV~\cite{mm}.

\begin{figure}
\begin{tabular}{cc}
  \includegraphics[height=.21\textheight]{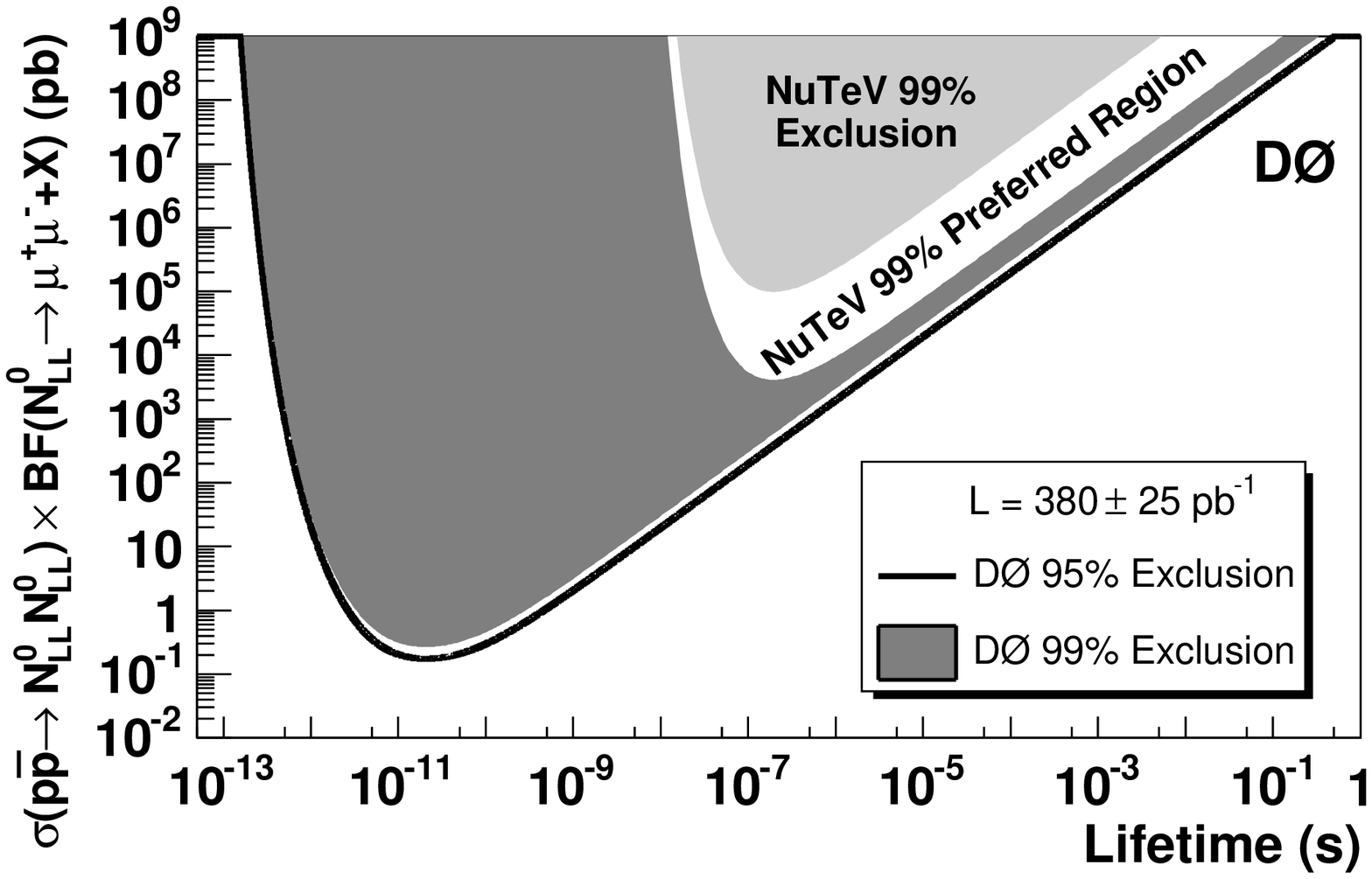} &
  \includegraphics[height=.23\textheight]{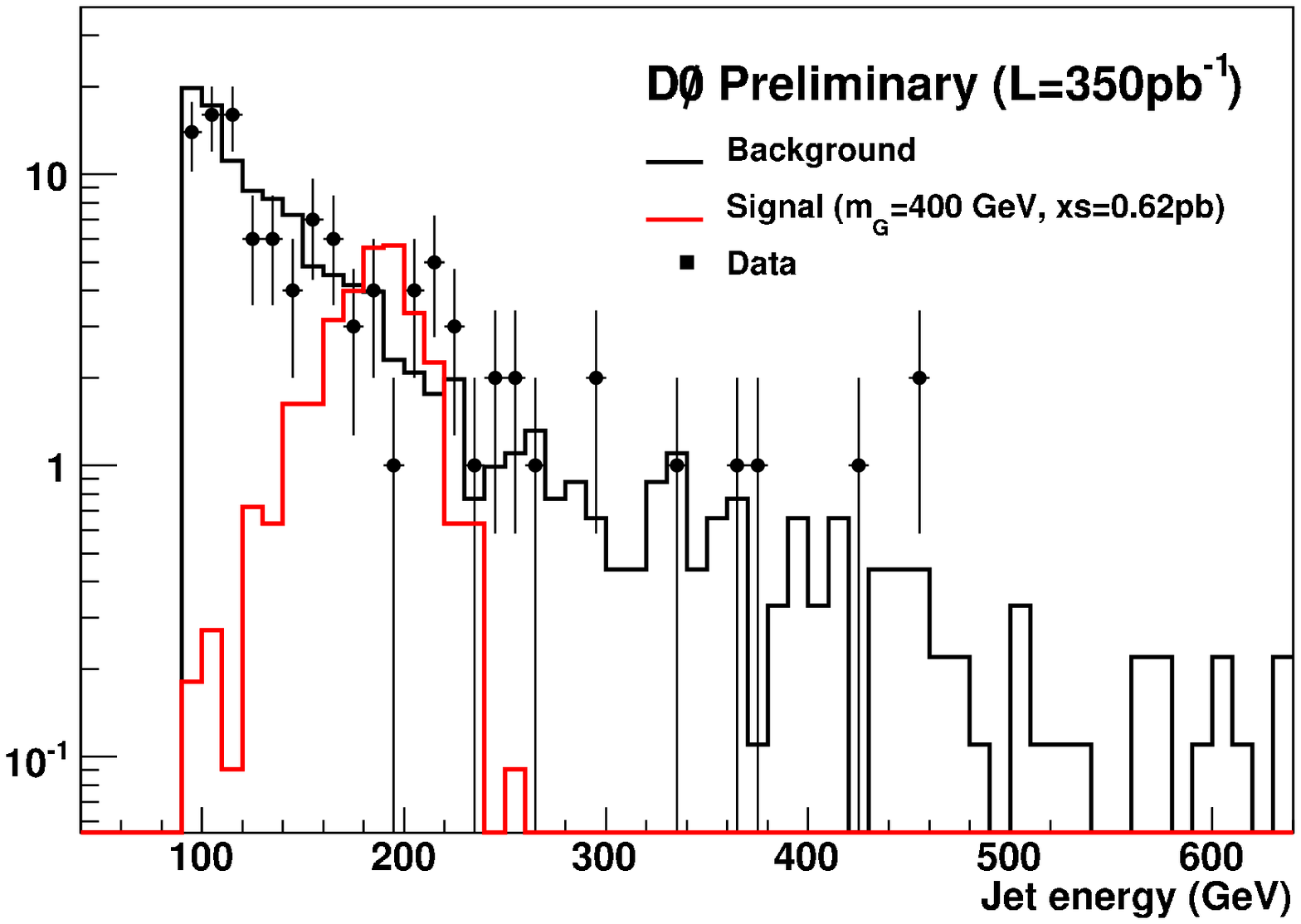}
  \caption{Upper limit on the production cross section times branching fraction
into two muons for neutral long lived particles, as a function of the lifetime 
(left). Jet energy spectrum in the search for long-lived gluinos (right).} 
\end{tabular}
  \label{nllp}
\end{figure}

Finally, a search for neutral long lived particles decaying 
into two muons and a neutrino via a small $\lambda_{122}$ coupling
was carried out by D\O~\cite{ca}. This search
was motivated by three anomalous events in this topology observed by the NuTeV 
collaboration~\cite{nutev}. 
The D\O\ analysis selected 0 events with a dimuon vertex well
displaced from the interaction point, while a background of $0.8 \pm 1.6$
events was expected. As shown in Fig.~\ref{nllp}, such an interpretation of
the NuTeV anomaly is excluded.

\subsection{GMSB, AMSB and Split-SUSY~\cite{yg}}

In the framework of gauge mediated SUSY breaking, the NLSP is expected to be
a stau or a neutralino. The latter case, where the NLSP decays into a photon 
and an invisible gravitino, was considered by D\O. Gaugino pair production 
ultimately leads to a final state containing two photons and missing $E_T$.
The backgrounds from electrons or jets misindentified as photons or from fake
missing $E_T$ due to jet energy mismeasurements were all evaluated from data,
leading to an expectation of $1.8 \pm 0.7$ events in 760~pb$^{-1}$, while four 
events were observed.
The $\tilde\chi^0_1$ mass lower limit thus obtained is 120~GeV, improving on the result
of a previous combination of CDF and D\O\ searches. 

Anomaly mediated SUSY breaking predicts a wino LSP almost degenerate in mass
with the lightest chargino, which may therefore well be long lived. Such a
long lived charged massive particle would appear in a collider detector as a
slowly moving muon. The timing information of the muon detector was used by
D\O\ to search for the pair production of such particles, and no candidate
events were found in 390~pb$^{-1}$ 
while a background of $0.69 \pm 0.05$ events was expected.
A chargino mass lower limit of 174~GeV was inferred.

In split SUSY, a variant which recently received much attention, all scalars
are very heavy, so that the gluinos become long lived. Such gluinos will
hadronize into $R$-hadrons and may come to rest in the D\O\ calorimeter. After
a while, they decay into a gluon and a $\tilde\chi^0_1$. The final state therefore 
appears as a jet with a random orientation in an otherwise empty event. The
backgrounds from cosmic or beam-related muons were evaluated from data, and
the predicted jet
energy spectrum was found to be well consistent with observation
(Fig.~\ref{nllp}). For $m_{\tilde\chi^0_1}<100$~GeV, a gluino mass lower limit of 
$300$~GeV was deduced.

\subsection{Indirect limits}

The $B_s\to\mu^+\mu^-$ decay is strongly suppressed in the SM, at the level of 
3 to 4~$10^{-9}$. A large enhancement is however expected in SUSY at large 
$\tan\beta$, proportional to $\tan^6\beta/m_A^4$. To overcome the huge 
background from SM dimuon production, CDF used a discriminant based on the
decay length between the primary and dimuon vertices, the isolation of the 
dimuon system, and the pointing of that system toward the primary vertex. 
One candidate event was 
selected in 780~pb$^{-1}$, while $1.4 \pm 0.4$ background events were expected.
The normalization was provided by the well measured branching fraction of 
$B^+\to J/\psi K^+$, with $J/\psi\to\mu^+\mu^-$. 
A 95\% C.L. upper limit of $10\times 10^{-8}$ was obtained~\cite{pm}, 
which already provides significant constraints on
SUSY at large $\tan\beta$~\cite{mc}.

\section{Extra dimensions}

There are now many variants of models with additional space dimensions. 
Two such models have received particular attention at the Tevatron. The 
Arkani-Hamed, Dimopololos and Dvali model~\cite{add} involves a number of large
extra-dimensions (EDs). 
The Randall and Sundrum (RS) model~\cite{rsth} calls for a 
single ED with a warped metric. In both of these models, only
gravity propagates in the EDs, and hence only the graviton has
Kaluza-Klein (KK) excitations.

The presence of large EDs can be probed in two ways:
\begin{itemize}
\item real KK gravitons are produced in association with a jet, the large 
number of kinematically accessible KK gravitons compensating for the smallness 
of the gravitational coupling. Here the final state topology is a monojet;
\item virtual KK gravitons are exchanged in the production of fermion or
vector boson pairs, thus interfering with SM production and modifying the
observed cross sections.
\end{itemize}

\begin{figure}
\begin{tabular}{cc}
  \includegraphics[height=.24\textheight]{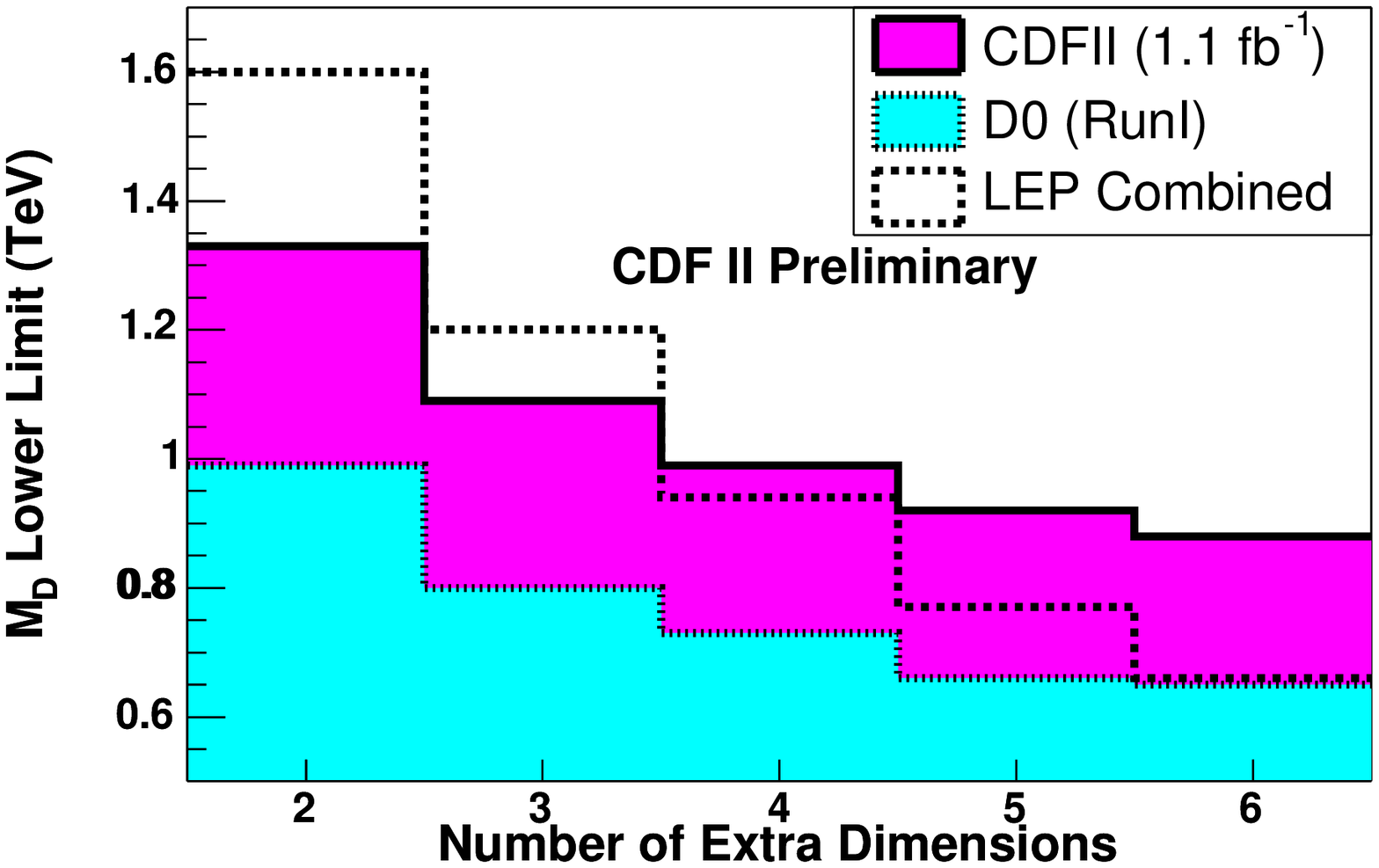} &
  \includegraphics[height=.24\textheight]{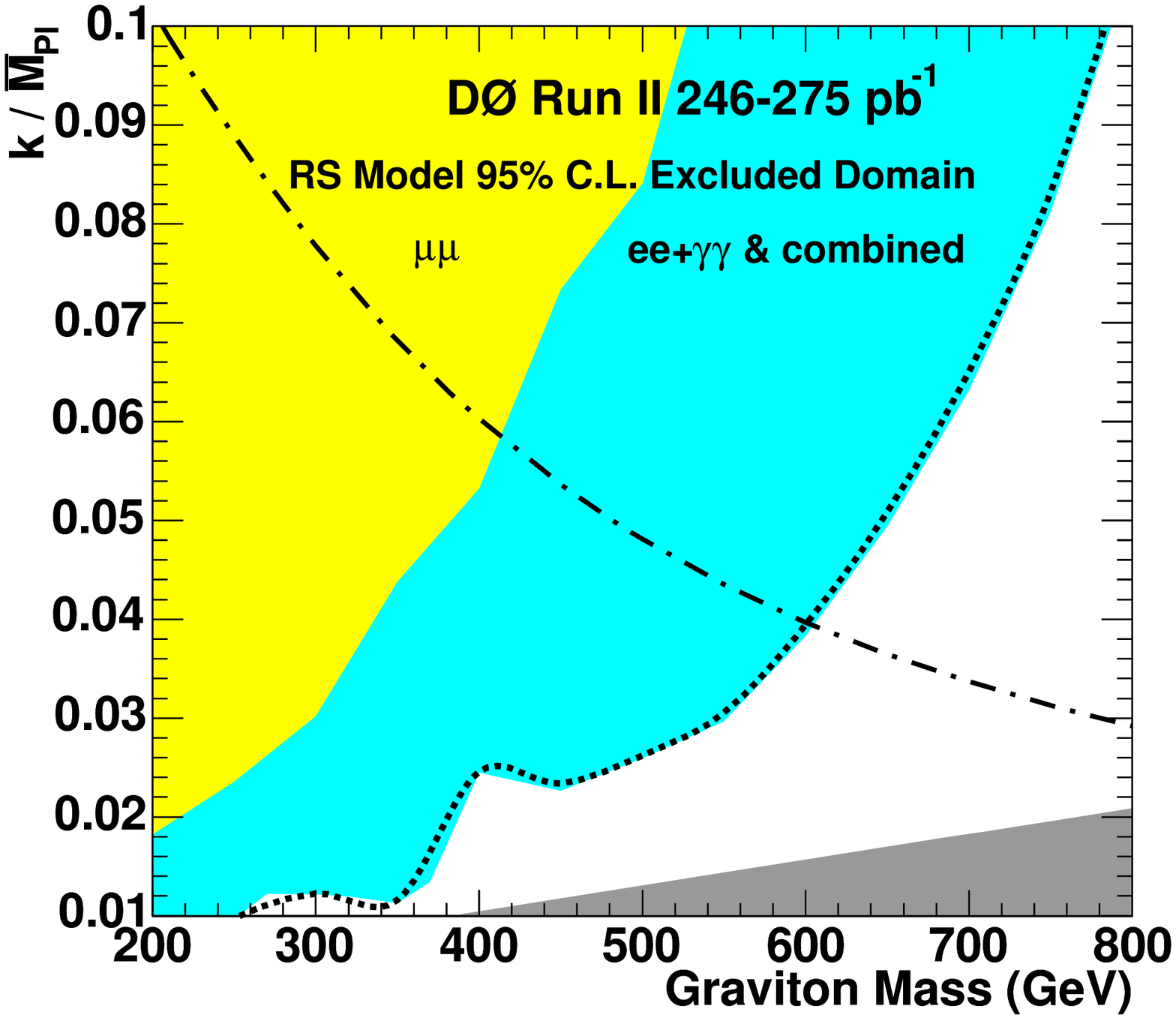}
  \caption{Lower limit on the fundamental Planck scale as a function of the 
number of large extra dimensions (left). Lower limit on the mass of an RS 
graviton as a function of its coupling to SM fields (right).} 
\end{tabular}
  \label{led}
\end{figure}

The CDF collaboration updated their search for monojets with 1.1~fb$^{-1}$. A
jet with $E_T>150$~GeV was required, allowing for soft additional jets. The
missing $E_T$, in excess of $120$~GeV, was required not to point along any jet
direction, and a veto on isolated leptons was required. The predicted 
background, dominated by $(Z\to\nu\nu)+$jet, amounted to $819 \pm 71$ events, 
while $779$ events were selected. Limits on the fundamental Planck scale $M_D$
were derived as a function of the number of EDs as shown in 
Fig.~\ref{led}~\cite{phb}.

To search for virtual effects from large EDs, D\O\ combined the dielectron and
diphoton final states and looked for deviations of the SM prediction at large
masses, combining the information from the mass spectrum with the angular 
distribution of the electrons and photons. Their results, based on 
200~pb$^{-1}$ and expressed in terms of an effective cutoff $M_S$ expected to 
be closely related to $M_D$, remain the most constraining to date. In the
formalism of Giudice, Ratazzi and Wells~\cite{grw}, 
they obtain $M_S>1.43$~TeV after combining with the D\O\ Run I result.

In the case of the RS model, the KK excitations have spacings of the order of
a TeV, and can be observed as dielectron, dimuon or diphoton resonances. 
Combining those channels,  
the D\O\ collaboration has set a mass lower limit on the first RS KK 
graviton as a function of its coupling $\kappa/M_\mathrm{Pl}$ to the SM 
fields (Fig.~\ref{led}). 

\section{Model independent searches}

The CDF collaboration performed a series of ``signature-based'' searches, where
possible departures from the SM predictions are looked for, irrespective of any
specific theoretical prejudice~\cite{mg}. Examples of such searches are 
analyses of the channels $\gamma\gamma+$X, where X is an electron, a muon or a
photon, or $\ell\gamma+$missing $E_T$ ($\ell= e$ or $\mu$), the latter being
motivated by an excess observed in the Run I data. All numbers of events and
kinematic distributions were found to be in agreement with the SM expectations.

\section{Final remarks}

The CDF and D\O\ collaborations have now begun to explore truly virgin 
territories. By the time of this conference, already 1.4~fb$^{-1}$ had been
recorded by each experiment, only partially deciphered. With a steadily 
improving performance of the Tevatron collider, there are still a number of 
years and fb$^{-1}$ of frontier physics ahead of us. Let's keep confident 
that results other than 95\% C.L. limits will be presented at one of the 
forthcoming SUSY conferences. 

\begin{theacknowledgments}
I wish to thank and congratulate Jonathan Feng and 
the organizers of the SUSY06 conference for
a remarkable (but somewhat exhausting\dots) scientific and social program. The 
help of the CDF exotic conveners Jane Nachtman and Song Ming Wang, as well as 
of my D\O\ colleague Arnd Meyer are greatfully acknowledged. 
\end{theacknowledgments}

\bibliographystyle{aipprocl} 

\begin{thebibliography}{9}

\bibitem{url} \url{http://www-cdf/fnal.gov/physics/exotic/exotic.html}\\
\url{http://www-d0.fnal.gov/Run2Physics/WWW/results/np.htm}.

\bibitem{phb} P.-H. Beauchemin, these proceedings.
\bibitem{js} J. Strologas, these proceedings.
\bibitem{mh} M. Hohlfeld, these proceedings.
\bibitem{pb} P. Bargassa, these proceedings.
\bibitem{mm} M. Martinez, these proceedings.
\bibitem{ca} C. Autermann, these proceedings.
\bibitem{nutev} T. Adams {\sl et al.}, 
Phys. Rev. Lett. {\bf 87}, 041801 (2001).
\bibitem{yg} Y. Gershtein, these proceedings.
\bibitem{pm} P. Maksimovic, these proceedings.
\bibitem{mc} M. Carena, these proceedings.
\bibitem{add} N. Arkani-Hamed, S. Dimopoulos and G. Dvali,
Phys. Lett. B {\bf 429}, 263 (1998).
\bibitem{rsth} L. Randall and R. Sundrum,
Phys. Rev. Lett. {\bf 83}, 3370 (1999).
\bibitem{grw} G. Giudice, R. Rattazzi and J. Wells,
Nucl. Phys. {\bf B544}, 3 (1999).
\bibitem{mg} M. Goncharov, these proceedings.

\end{thebibliography}

\end{document}